\documentclass[12pt, epsfig]{article}                            
\usepackage{epsfig}       
\usepackage{amssymb}     
\usepackage{amsfonts}     

\newskip\humongous \humongous=0pt plus 1000pt minus 1000pt

\newif\ifdtup

\catcode`\@=11
 
\@addtoreset{equation}{section}
\def\theequation{\thesection.\arabic{equation}}
 
\def\@normalsize{\@setsize\normalsize{15pt}\xiipt\@xiipt
\abovedisplayskip 14pt plus3pt minus3pt%
\belowdisplayskip \abovedisplayskip
\abovedisplayshortskip \z@ plus3pt%
\belowdisplayshortskip 7pt plus3.5pt minus0pt}
 
\def\small{\@setsize\small{13.6pt}\xipt\@xipt
\abovedisplayskip 13pt plus3pt minus3pt%
\belowdisplayskip \abovedisplayskip
\abovedisplayshortskip \z@ plus3pt%
\belowdisplayshortskip 7pt plus3.5pt minus0pt
\def\@listi{\parsep 4.5pt plus 2pt minus 1pt
     \itemsep \parsep
     \topsep 9pt plus 3pt minus 3pt}}
 
\relax

\catcode`@=12
 
\catcode`\@=11
 
\def\section{\@startsection{section}{1}{\z@}{3.5ex plus 1ex minus
   .2ex}{2.3ex plus .2ex}{\large\bf}}
 
\def\thesection{\arabic{section}}    
\def\thesubsection{\arabic{section}.\arabic{subsection}}

\def\theequation{\arabic{section}.\arabic{equation}}
\def\appendix{\setcounter{section}{0}
 \def\thesection{Appendix \Alph{section}}
 \def\thesubsection{\Alph{section}.\arabic{subsection}}
 \def\theequation{\Alph{section}.\arabic{equation}}}
 
\def\YGrule{0.4}   
\def\YGbox{6.5}    
\def\SymBoxes#1#2#3#4{\newdimen\un@t \un@t#3%
\raisebox{#1}{\rule{#2\un@t}{#4}\hskip-#2\un@t
\@tempdimb\un@t \advance\@tempdimb by-#4\@tempcntb#2\relax%
\@whilenum{\@tempcntb>0}\do{
\rule{#4}{\un@t}\hskip\@tempdimb \advance\@tempcntb by\m@ne}%
\hskip-#2\un@t \rule[\un@t]{#2\un@t}{#4}%
\rule[\un@t]{#4}{#4}\hskip-#4
\rule{#4}{\un@t}}\hskip-#4}                
\def\Young{\@ifnextchar[{\@Young}{\@Young[0]}}
\def\@Young[#1]#2{\newdimen\YG@unit \YG@unit\YGbox pt%
\newdimen\h@ight \h@ight#1\YG@unit \@tempcnta-1\relax
\@tfor\c@ount:=#2\do{\advance\@tempcnta by\@ne}
\@tempdima\@tempcnta\YG@unit%
\advance\h@ight by\@tempdima\relax     
\@tfor\c@ount:=#2\do{\SymBoxes{\h@ight}{\c@ount}{\YG@unit}{\YGrule pt}%
\@tempdima-\c@ount\YG@unit \hskip\@tempdima%
\advance \h@ight by -\YG@unit}         
\@tempdima\YG@unit \multiply\@tempdima by\@car#2\@nil %
\hskip\@tempdima}                      
\def\YoungTab{\@ifnextchar[{\@YoungIdx}{\@YoungIdx[0]}}
\def\@YoungIdx[#1]{\@ifnextchar[{\@iYoungIdx[#1]}{\@iYoungIdx[#1][\@empty]}}
\def\@iYoungIdx[#1][#2]#3{%
\newdimen\YG@unit \YG@unit\YGbox pt\newdimen\YG@rule \YG@rule \YGrule pt
\newcount\c@ount \c@ount\z@ \newdimen\skip@wd \unitlength\@ne pt
\newdimen\h@ight \h@ight#1\YG@unit \@tempcnta\m@ne\relax
\@tfor\d@um:=#3\do{\advance\@tempcnta by\@ne}
\@tempdima\@tempcnta\YG@unit%
\advance\h@ight by\@tempdima\relax
\@tfor\@idxlist:=#3\do{
\@tempcnta\z@\hskip.5\YG@rule\relax 
\@for\@idx:=\@idxlist\do{
\raisebox{\h@ight}{\makebox(\YGbox,\YGbox){#2$\@idx$}}
\advance\@tempcnta by\@ne}\hskip-.5\YG@rule%
\@tempdima-\@tempcnta\YG@unit \hskip\@tempdima%
\ifnum\c@ount=\z@ \skip@wd-\@tempdima\fi \relax
\SymBoxes{\h@ight}{\@tempcnta}{\YG@unit}{\YG@rule}%
\hskip\@tempdima \advance\h@ight by -\YG@unit
\advance\c@ount by\@ne}
\hskip\skip@wd}                      
\def\p{{}^{\prime}}
\def\pp{{}^{\prime\prime}}
\newcommand{\piup}{\hspace{-.05cm}+\hspace{-.05cm}}
\newcommand{\menop}{\hspace{-.05cm}-\hspace{-.05cm}}
\newcommand{\n}{{\mathbf{n}}}
\newcommand{\np}{{\mathbf{n\p}}}   
\newcommand{\nd}{{\mathbf{\dot{n}}}}   
\newcommand{\npn}{{(\mathbf{n\p\times n})}}   
\newcommand{\ndn}{{(\mathbf{\dot{n}\times n})}}   
\newcommand{\F}{{\mathbf{F}}}   
\newcommand{\Q}{{\mathbf{Q}}}   
\newcommand{\J}{{\mathbf{J}}}   
\newcommand{\A}{{\mathbf{A}}}   
\newcommand{\beq}{\begin{equation}}
\newcommand{\eeq}{\end{equation}}

\begin{document}

\newcommand{\be}{\begin{equation}}
\newcommand{\ee}{\end{equation}}
\newcommand{\bqa}{\begin{eqnarray}}
\newcommand{\bea}{\begin{eqnarray}}
\newcommand{\eea}{\end{eqnarray}}
\newcommand{\beas}{\begin{eqnarray*}}
\newcommand{\eeas}{\end{eqnarray*}}
\newcommand{\defi}{\stackrel{\rm def}{=}}
\newcommand{\non}{\nonumber}
\newcommand{\bquo}{\begin{quote}}
\newcommand{\enqu}{\end{quote}}
\def\de{\partial}
\def\Tr{ \hbox{\rm Tr}}
\def\tr{ \hbox{\rm tr}}
\def\const{\hbox {\rm const.}}
\def\o{\over}
\def\im{\hbox{\rm Im}}
\def\re{\hbox{\rm Re}}
\def\ket#1{|{#1}\rangle}
\def\bra#1{\langle {#1} |}
\def\ckt{\rangle}
\def\brc{\langle}
\def\hsp{,\hspace{.5cm}}
\def\hspp{,\hspace{.3cm}}

\def\Arg{\hbox {\rm Arg}}
\def\Re{\hbox {\rm Re}}
\def\Im{\hbox {\rm Im}}
\def\diag{\hbox{\rm diag}}
\def\longvert{{\rule[-2mm]{0.1mm}{7mm}}\,}

{\hfill     IFUP-TH/2010-35} 
\bigskip

\begin{center}
{\large  {\bf  
A Faddeev-Niemi Solution that Does Not Satisfy Gauss' Law
 } } 
\end{center}

\bigskip
\begin{center}
{\large  Jarah Evslin$^{1,3}$ and Simone Giacomelli$^{2,3}$   \vskip 0.10cm
 }
\end{center}

\vspace{1cm}

\begin{center}
{\it   \footnotesize
Dipartimento di Fisica ``E. Fermi'' -- Universit\`a di Pisa $^{(1)}$, \\
   Largo Bruno Pontecorvo, 3, Ed. C, 56127 Pisa,  Italy \\
Scuola Normale Superiore - Pisa $^{(2)}$,
 Piazza dei Cavalieri 7, Pisa, Italy \\
Istituto Nazionale di Fisica Nucleare -- Sezione di Pisa $^{(3)}$, \\
     Largo Bruno Pontecorvo, 3, Ed. C, 56127 Pisa,  Italy 
   }

\end {center}

\vspace{2cm}

\noindent  
{\bf Abstract:}
Faddeev and Niemi have proposed a reformulation of SU(2) Yang-Mills theory in terms of a U(1) gauge theory with 8 off-shell degrees of freedom.  We present a solution to Faddeev and Niemi's formulation which does not solve the SU(2) Yang-Mills Gauss constraints.  This demonstrates that the proposed reformulation is inequivalent to Yang-Mills, but instead describes Yang-Mills coupled to a particular choice of external charge.
\vfill  
 
\begin{flushright}
  \today
\end{flushright}

\bigskip

\hfill{}

\section{Introduction}

A classical SU(2) Yang-Mills field configuration is characterized by a connection $\A_\mu$, a triplet of 4-vectors which therefore contains 12 off-shell degrees of freedom.   On-shell there are less.  3 of the equations of motion, those obtained by varying the potential $\A_t$, are first order in time derivatives.  Therefore these constrain the Cauchy data, the values of the fields and their first derivatives on an initial surface, leaving only 9 degrees of freedom.  Furthermore a finite propagator requires an invertible kinetic term.  The kinetic term has a three-dimensional kernel, and so it becomes invertible only when a further 3 degrees of freedom are (gauge) fixed, leaving 6 on-shell degrees of freedom.

Thus the usual description of Yang-Mills theory involves twice as many degrees of freedom as actually propagate.  As in the ADM formulation of gravity \cite{ADM}, one may solve the constraints and fix the gauge, but at a price of losing manifest Lorentz covariance.  In Refs. \cite{Shabanov,Kondo} the authors partially solved the constraints and fixed the gauge invariance while preserving manifest Lorentz covariance, however the price is that their solution is not written explicitly, but rather is defined implicitly in terms of the solution of an extremization problem.  

Faddeev and Niemi have claimed \cite{FN} that one may partially fix the gauge and solve the constraints explicitly in a way that preserves manifest Lorentz-invariance.  They have proposed the following decomposition of the $SU(2)$ Yang-Mills connection $\A_\mu$ 
\beq
\A_\mu=C_\mu\n +\partial_\mu \n\times\n +\rho\partial_\mu\n+\sigma\partial_\mu \n\times\n \label{FN}
\eeq
where $C_\mu$ is a $U(1)$ connection, $\sigma$ and $\rho$ are scalars and $\n$ is a unit vector in the $\mathfrak{su}(2)$ Lie algebra.  This contains 8 off-shell degrees of freedom, one of which is pure gauge and one of which is eliminated by the Gauss constraint for the $U(1)$ gauge field, leaving 6 degrees of freedom on-shell as in Yang-Mills theory.   

As all choices of the 8 variables $C_\mu ,\ \rho,\ \sigma$ and $\n$ lead to values of $\A_\mu$, the variational principle implies that the Faddeev-Niemi equations of motion will be a 7-dimensional subset of the 9-dimensional Yang-Mills equations of motion, where we have used the fact that variations in the pure gauge directions (1 for Faddeev-Niemi, 3 for Yang-Mills) yield trivial equations of motion.  Therefore (\ref{FN}) is equivalent to the original Yang-Mills theory if the missing two equations of motion, which may be taken to be first order (constraints) by adding suitable second order equations of motion, are automatically solved by the decomposition (\ref{FN}).  In other words, if one may derive (\ref{FN}) by imposing two Gauss constraints on a general choice of SU(2) connection and choosing a gauge, then the Faddeev-Niemi decomposition is equivalent to the original Yang-Mills theory. 

In this note we will give an example of a solution of the Faddeev-Niemi equations of motion which does not satisfy the Yang-Mills Gauss' law, demonstrating that (\ref{FN}) is not equivalent to Yang-Mills.  However we will argue that it does satisfy the Yang-Mills equations with a different Gauss' law, corresponding to an external charge.  It therefore describes a truncation of QCD where instead of eliminating the quarks as in Yang-Mills theory, one considers nonpropagating quarks which yield a gluon-dependent charge density.

\section{The Faddeev-Niemi equations of motion}

\subsection{Deriving the equations}

The SU(2) gauge field strength (we use the convention $\nabla_{\mu}=\partial_{\mu}-\imath[A_{\mu},\;]$, so $F_{\mu\nu}=\partial_{\mu}A_{\nu}-\partial_{\nu}A_{\mu}-\imath [A_{\mu},A_{\nu}]$) obtained from the connection (\ref{FN}) is
\beq
\F_{\mu\nu}=(G_{\mu\nu}-[1-(\rho^2+\sigma^2)]H_{\mu\nu})\n+D_{[\mu}\rho\partial_{\nu]}\n+D_{[\mu}\sigma\partial_{\nu]}\n\times\n \label{F}
\eeq
where square brackets denote antisymmetrization with no factor of $1/2$ and $D$ is a covariant derivative using the $U(1)$ connection $C$
\beq
D_{\mu}\rho=\partial_{\mu}\rho+C_{\mu}\sigma,\qquad D_{\mu}\sigma=\partial_{\mu}\sigma-C_{\mu}\rho
\eeq
 and
\beq
G_{\mu\nu}=\partial_{[\mu}C_{\nu]}\hsp
H_{\mu\nu}=\n\cdot (\partial_\mu\n\times\partial_\nu\n).
\eeq

The 8 off-shell degrees of freedom yield the following 8 equations
\bea
\n\cdot\nabla^\mu\F_{\mu\nu}&=&0\label{con}\\
\partial^\nu\n\cdot\nabla^\mu\F_{\mu\nu}&=&0\label{5}\\
\partial^\nu\n\times\n\cdot\nabla^\mu\F_{\mu\nu}&=&0\label{6}\\
(D^\nu\rho+D^\nu\sigma \n\times)\nabla^\mu\F_{\mu\nu}&=&0\label{7}
\eea
where $\nabla$ is the the $SU(2)$-covariant derivative.  Eq.~(\ref{con}) contains 4 equations, Eq.~(\ref{5}) and Eq.~(\ref{6}) each contain one and finally Eq.~(\ref{7}) contains two independent equations.  

\subsection{The constraints, charges and currents}

Notice that the three SU(2) Gauss constraints
\beq
\nabla^\mu\F_{\mu t}=0
\eeq
are not obviously included in the equations of motion, which are all various projections of the SU(2) equations.  One of the three constraints, that in the direction $\n$ of the $U(1)$, is however contained in the $\nu=t$ case of Eq.~(\ref{con}).  

To interpret the equations of motion, we will consider a simplified ansatz in which the $y$ and $z$ components of the $U(1)$ connection $C_{\mu}$ vanish and all fields are constant in the $y$ and $z$ directions.  Therefore the only nontrivial component of the $SU(2)$ field strength is $F_{tx}=-F_{xt}$.  In this case the $\mathfrak{su}(2)$-valued charge is simply
\beq
\Q_{\rm{tot}}=-\partial_\mu \F_{\mu t}=\partial_x \F_{tx}. 
\eeq
As in any nonabelian gauge theory, this charge is generically nonzero, as the gluons themselves are charged.  Indeed in pure Yang-Mills the field strength is not conserved, but is rather only covariantly conserved.  Any violation of the covariant conservation corresponds to an external charge
\beq
\Q=\nabla_x F_{tx}=\partial_x \F_{tx}-\imath[\A_x,\F_{tx}]. \label{Q}
\eeq
In pure Yang-Mills, the Gauss constraint which follows from the variation of the temporal components of the gauge field sets $\Q=0$.  In the Faddeev-Niemi decomposition, not all of the temporal components of the gauge field are manifestly independent, and therefore the vanishing of $\Q$ is not automatic and below we will see an example in which $\Q$ does not vanish.

Similarly we may define an SU(2) current.  The $y$ and $z$-independent ansatz guarantees that the $y$ and $z$ components of the current vanish, thus there is only a nontrivial current in the $x$ direction
\beq
\J_{\rm{tot}}=\partial_\mu \F_{\mu x}=\partial_t \F_{tx}.
\eeq
As in our ansatz $\F$ contains no components with two spatial indices, and we are working in flat Minkowski space, the fact that we have not raised the indices on the derivative merely causes a sign in the definition of $\J$ and has no physical consequences.  Again this current may be decomposed into a current consisting of gluons and an external current
\beq
\J=\partial_t \F_{tx}-\imath[\A_t,\F_{tx}]. \label{J}
\eeq

Both the external charge $\Q$ and the external current $\J$ are $\mathfrak{su}(2)$-valued, thus they each contain 3 components.  It will be convenient to choose two bases in which to expand them.  Recall that $\n$ is a unit vector in $\mathfrak{su}(2)$.  Therefore $\np=\partial_x\n$ and $\npn$ are both orthogonal to $\n$ and to each other.  Together with $\n$ they thus provide an orthogonal but in general not orthonormal basis.  Another orthogonal basis is provided by $\n$, $\nd=\partial_t\n$ and $\ndn$. Of course, when any of these derivatives vanish, the corresponding triplet no longer spans the $\mathfrak{su}(2)$ Lie algebra. The decompositions in these bases are simply
\bea
&&Q^0=\n\cdot\Q\hsp
Q^1=\np\cdot\Q\hsp
Q^2=\npn\cdot\Q\\
&&J^0=\n\cdot\J\hsp
J^1=\np\cdot\J\hsp
J^2=\npn\cdot\J\nonumber\\
&& \tilde{Q}^1=\nd\cdot\Q\hspp
\tilde{Q}^2=\ndn\cdot\Q\hspp
\tilde{J}^1=\nd\cdot\J\hspp
\tilde{J}^2=\ndn\cdot\J\nonumber.
\eea

\subsection{The equations of motion in terms of charges and currents}

The equations of motion (\ref{con}), (\ref{5}) and (\ref{6}) are easily reexpressed in terms of these charges and currents.  Eq.~(\ref{con}) contains four equations, one for each value of the index $\nu$.  The $\nu=y$ and $\nu=z$ equations are trivial, while the $\nu=t$ equation is the $U(1)$ Gauss constraint
\beq
Q^0=0
\eeq
and $\nu=x$ states that the $U(1)$ current also vanishes
\beq
J^0=0.
\eeq
Equations (\ref{5}) and (\ref{6}) are simply
\beq
\tilde{Q}^1=J^1\hsp
\tilde{Q}^2=J^2. \label{ohm}
\eeq
The last equation, Eq.~(\ref{7}), is a vector in $\mathfrak{su}(2)$ and so consists of 3 equations.  As the U(1) component of the charge and current vanishes, only two of these equations are independent as expected since the 8 off-shell degrees of freedom of the Faddeev-Niemi decomposition should lead to 8 equations of motion. In terms of the currents and charges they can be written in the form
\bea
&&(\dot{\rho}+C_t\sigma)Q^1+(\dot{\sigma}-C_t\rho)Q^2=(\rho^{\prime}+C_x\sigma)J^1+(\sigma^{\prime}-C_x\rho)J^2,\\
&&(\dot{\rho}+C_t\sigma)Q^2-(\dot{\sigma}-C_t\rho)Q^1=(\rho^{\prime}+C_x\sigma)J^2-(\sigma^{\prime}-C_x\rho)J^1.\nonumber
\eea

The equations (\ref{ohm}) relating the current and charge may appear quite strange.  However, the change of coordinates has units of velocity, and so this is the familiar equation relating the current to the charge times the velocity.  We may therefore use it to define the velocity of the external charge carriers.
To transform between the two bases one needs the identities
\beq
\np=\frac{\nd\cdot\np}{|\nd|^2}\nd+\frac{\ndn\cdot\np}{|\nd|^2}\ndn\hsp
\npn=\frac{\nd\cdot\npn}{|\nd|^2}\nd+\frac{\nd\cdot\np}{|\nd|^2}\ndn.
\eeq
One may then expand $\Q$ as follows
\bea
\Q&=&Q^0\n+\frac{Q^1}{|\np|^2}\np+\frac{Q^2}{|\np|^2}\npn\\
&=&Q^0\n\hspace{-.00cm}+\hspace{-.00cm}\frac{Q^1(\np\cdot\nd)}{|\np|^2|\nd|^2}\nd\hspace{-.00cm}+\hspace{-.00cm}\frac{Q^1(\np\cdot\ndn)}{|\np|^2|\nd|^2}\ndn\nonumber\\
&&\hspace{-.00cm}+\hspace{-.00cm}\frac{Q^2(\npn\cdot\nd)}{|\np|^2|\nd|^2}\nd\hspace{-.00cm}+\hspace{-.00cm}\frac{Q^2(\np\cdot\nd)}{|\np|^2|\nd|^2}\ndn.\nonumber
\eea
Identifying the coefficients of $\nd$ and $\ndn$ with $\tilde{Q}^1$ and $\tilde{Q}^2$ respectively one may rewrite the equations of motion (\ref{ohm}) as
\beq
\left(
\begin{array}{c}
J^1\\J^2
\end{array}
\right)
=
\left(
\begin{array}{c}
\tilde{Q}^1\\
\tilde{Q}^2
\end{array}
\right)
=
\frac{1}{|\np|^2}
\left(
\begin{array}{cc}
\nd\cdot\np&\nd\cdot\npn\\
-\nd\cdot\npn&\nd\cdot\np
\end{array}
\right)
\left(
\begin{array}{c}
Q^1\\
Q^2
\end{array}
\right) . 
\eeq
For example, if at a given point we boost in the $x$-direction to arrive at a reference frame in which $\nd$ and $\np$ are parallel, then this becomes
\beq
J^k=\frac{\nd\cdot\np}{|\np|^2}Q^k=\frac{|\nd||\np|}{|\np|^2}Q^k=\frac{|\nd|}{|\np|}Q^k.
\eeq
This identifies the velocity of the external charge in the direction $k$ with the phase velocity of $\n$ in the direction $k$.  

In the Faddeev-Niemi decomposition, $\n$ is notoriously hard to interpret.  It is only in very particular configurations, such as magnetic monopoles that it yields an unbroken $U(1)$ gauge symmetry.  This formula therefore yields an interpretation, the external charges are in some sense bound to the phase of $\n$, as plasma may be bound to a magnetic flux tube.  The identification of the velocity of a physical charge with a phase velocity, which may be faster than the speed of light, may imply a breakdown of causality in this system.

\section{A formula for the external charges and currents}

This discussion would be quite irrelevant if the Faddeev-Niemi equations of motion implied that $\Q=\J=0$.  We will now calculate $\Q$ and $\J$ in this ansatz, demonstrating that the equations of motion may be solved by a configuration with nonvanishing external $SU(2)$ charge and current.  The Faddeev-Niemi decomposition in our ansatz immediately yields the SU(2) connection
\beq
\A_x=C_x\n+\rho\np+(1+\sigma)\npn\hsp
\A_t=C_t\n+\rho\nd+(1+\sigma)\ndn.
\eeq
Similarly the $SU(2)$ field strength is given by Eq.~(\ref{F})
\beq
F_{tx}=[G_{tx}+(1-\rho^2-\sigma^2)\alpha]\n+D_t\rho\np-D_x\rho\nd+D_t\sigma\npn-D_x\sigma\ndn
\eeq
where we have defined the function
\beq
\alpha=\n\cdot(\np\times\nd).
\eeq
The external charge may then be computed directly from the definition (\ref{Q})
\bea\label{q}
\Q\hspace{-.3cm}&=&\hspace{-.3cm}[G_{tx}^{\prime}-3(\rho\rho\p\piup\sigma\sigma\p)\alpha\piup(1-\rho^2-\sigma^2)\alpha\p+(\rho D_x\sigma\menop\sigma D_x\rho)\nd\cdot\np\piup(\sigma D_t\rho-\rho D_t\sigma)|\np|^2]\n\nonumber\\
&&+[C_xD_t\sigma+(D_t\rho)^{\prime}+\sigma G_{xt}+\sigma(\rho^2+\sigma^2-1)\alpha]\np+(C_xD_x\rho-(D_x\sigma)^{\prime})\ndn\nonumber\\
&&+[(D_t\sigma)^{\prime}-C_{x}D_t\rho-\rho G_{xt}-\rho(\rho^2+\sigma^2-1)\alpha]\npn-((D_x\rho)^{\prime}+C_xD_x\sigma)\nd\nonumber\\
&&-D_x\rho\dot\n\p_\perp+D_t\rho\n\pp_\perp
+D_t\sigma\n\pp\times\n-D_x\sigma(\dot\np\times\n).
\eea
In the previous equation we have defined the components orthogonal to $\n$ of $\dot\n\p$ and $\n\pp$ as $\dot\n\p_\perp$ and $\n\pp_\perp$ respectively. Similarly (\ref{J}) yields the external current
\bea
\J\hspace{-.3cm}&=&\hspace{-.3cm}[\dot{G_{tx}}-3(\rho\dot\rho\piup\sigma\dot\sigma)\alpha\piup(1-\rho^2-\sigma^2)\dot\alpha\piup(\rho D_x\sigma\menop\sigma D_x\rho)|\nd|^2\piup(\sigma D_t\rho\menop\rho D_t\sigma)(\np\cdot\nd)]\n\nonumber\\
&&-[C_tD_x\sigma+\dot{(D_x\rho)}+\sigma G_{tx}+\sigma(1-\rho^2-\sigma^2)\alpha]\nd-(C_tD_t\rho-\dot{D_t\sigma})\npn\nonumber\\
&&+[C_{t}D_x\rho-\dot{(D_x\sigma)}+\rho G_{tx}+\rho(1-\rho^2-\sigma^2)\alpha]\ndn+(\dot{D_t\rho}+C_tD_t\sigma)\np\nonumber\\
&&+D_t\rho\dot\n\p_\perp+D_t\sigma\dot\n\p\times\n-D_x\rho\ddot\n_\perp-D_x\sigma(\ddot\n\times\n).\label{j}
\eea

The equations of motion (\ref{con}) are the vanishing of the charge and current in the $\n$ direction
\bea
0\hspace{-.3cm}&=&\hspace{-.3cm}Q^0\hspace{-.1cm}=\hspace{-.1cm}G_{tx}^{\prime}-3(\rho\rho\p\piup\sigma\sigma\p)\alpha\piup(1-\rho^2-\sigma^2)\alpha\p+(\rho D_x\sigma\menop\sigma D_x\rho)\nd\cdot\np\nonumber\\
&&+(\sigma D_t\rho-\rho D_t\sigma)|\np|^2\\
0\hspace{-.3cm}&=&\hspace{-.3cm}J^0\hspace{-.1cm}=\hspace{-.1cm}\dot{G_{tx}}-3(\rho\dot\rho\piup\sigma\dot\sigma)\alpha\piup(1-\rho^2-\sigma^2)\dot\alpha\piup(\rho D_x\sigma\menop\sigma D_x\rho)|\nd|^2\nonumber\\
&&+(\sigma D_t\rho-\rho D_t\sigma)(\np\cdot\nd)\ .
\eea
The other Faddeev-Niemi equations can similarly be derived from Eqs.~(\ref{q}) and (\ref{j}). 
\section{A solution with nonzero external charge}

These equations are quite difficult to solve in general, but in order to find a solution which does not satisfy the Yang-Mills equations it is sufficient to consider the class of solutions with $\n$ independent of time. This assumption greatly simplifies the Faddeev-Niemi equations and one can see that in order to satisfy (\ref{con}), (\ref{5}) and (\ref{6}) it is sufficient to require $\J=Q^{0}=0$. Equation (\ref{7}) is now solved by imposing $D_t\rho=D_t\sigma=0$.
The equation $\J=0$ reads
\beq
\partial_{t}G_{tx}=0,\quad \ddot\rho-C_{t}^{2}\rho+\dot{C_t}\sigma+2C_t\dot{\sigma}=0,\quad \ddot\sigma-C_{t}^{2}\sigma-\dot{C_t}\rho-2C_t\dot{\rho}=0.
\eeq

Our solution of the Faddeev-Niemi equations is 
\beq\label{sol}
\begin{array}{ll}
C_t=0, & \rho=\rho_0,\nonumber\\
C_x=a+bt, & \sigma=\sigma_0 \label{noi}
\end{array}
\eeq
where $a$, $b$, $\rho_0$ and $\sigma_0$ are constants. Note that this form directly implies $D_t\rho=D_t\sigma=0$ and $Q^{0}=0$ regardless of the specific form of $\n$ (provided that it is independent of time).  The external currents $\J$ vanish, as do $\tilde{Q}^1$ and $\tilde{Q}^2$  since $\nd=0$.  

However our solution (\ref{sol}) does not satisfy Yang-Mills equations. By substituting (\ref{noi}) into (\ref{q}) and (\ref{j}) we find
\beq
\J=0,\qquad \Q=b\rho_0\npn-b\sigma_0\np.
\eeq 
Therefore these solutions violate the $SU(2)$ Gauss constraints, demonstrating that the Faddeev-Niemi parametrization (\ref{FN}) is not equivalent to Yang-Mills theory, but requires an external charge source.  Such an
 external charge may be obtained, for example, via an embedding of Yang-Mills into QCD.  However unlike QCD the dynamics of the charges, the values of $Q$ and $J$, are determined entirely by the parameters $\n,$ $C_\mu$, $\sigma$ and $\rho$ of the gauge field.  There are no propagating quark degrees of freedom.

Nonetheless the parametrization proposed by Faddeev and Niemi contains most of the on-shell degrees of freedom of the original Yang-Mills theory. First of all we can notice that the only ``sources of violation'' of the Yang-Mills equations are $\rho$ and $\sigma$.
Indeed, if we set $\rho=\sigma=0$ we are left with the Cho connection \cite{Cho}
\beq
A_{\mu}=C_{\mu}\n+\partial_\mu \n\times\n
\eeq
and by direct computation one can check that the only non trivial contributions to $\Q$ and $\J$ are directed along $\n$
so the equation (\ref{con}) implies all of the Yang-Mills equations. 

Therefore a subsector of the Faddeev-Niemi theory containing at least 4 of the 6 degrees of freedom is equivalent to a subsector of Yang-Mills theory.  This subsector contains several configurations of physical interest, such as the Wu-Yang monopole \cite{WuYang}.
Furthermore, there are also ``physical configurations'' with nontrivial $\rho$ and $\sigma$ fields: as noted by Faddeev and Niemi \cite{FN2}, the Ansatz
\beq
C_i=Cx^i,\qquad n^a=\frac{x^a}{r},\nonumber
\eeq
with $C$, $C_t$, $\rho$ and $\sigma$ depending on $t$ and $r$ only, coincides with the Witten's Ansatz for instantons \cite{Witten}. 

\section*{Acknowledgments}

\noindent
We are indebted to Sven Bjarke Gud\hspace{-.2cm}${}^-$\hspace{-.02cm}nason, Kenichi Konishi and Alberto Michelini for numerous discussions.



\begin{thebibliography}{23}

\bibitem{ADM}
 R.~L.~Arnowitt, S.~Deser and C.~W.~Misner,
  ``The dynamics of general relativity,''
  arXiv:gr-qc/0405109.

\bibitem{Shabanov}
 S.~V.~Shabanov,
  ``An effective action for monopoles and knot solitons in Yang-Mills
  theory,''
  [arXiv:hep-th/9903223].
 S.~V.~Shabanov,
  ``Yang-Mills theory as an Abelian theory without gauge fixing,''
  [arXiv:hep-th/9907182].
 
 
\bibitem{Kondo}
  K.~I.~Kondo, T.~Murakami and T.~Shinohara,
  ``Yang-Mills theory constructed from Cho-Faddeev-Niemi decomposition,''
  [arXiv:hep-th/0504107].
   K.~I.~Kondo, T.~Murakami and T.~Shinohara,
  ``BRST symmetry of SU(2) Yang-Mills theory in Cho-Faddeev-Niemi
  decomposition,''
  [arXiv:hep-th/0504198].
 
 
\bibitem{FN}
 L.~D.~Faddeev and A.~J.~Niemi,
  ``Partially dual variables in SU(2) Yang-Mills theory,''
  [arXiv:hep-th/9807069].
 
\bibitem{Cho}
 Y.~M.~Cho,
  ``A Restricted Gauge Theory,''
  Phys.\ Rev.\  D {\bf 21} (1980) 1080.
 Y.~M.~Cho,
  ``Extended Gauge Theory And Its Mass Spectrum,''
  Phys.\ Rev.\  D {\bf 23} (1981) 2415. 

\bibitem{WuYang}
 T.~T.~Wu and C.~N.~Yang,
  ``Some Solutions Of The Classical Isotopic Gauge Field Equations,''
in *H. Mark and S. Fernbach, Properties Of Matter Under Unusual Conditions*, New York 1969, 349-345. 

\bibitem{FN2}
 L.~D.~Faddeev and A.~J.~Niemi,
  ``Partial duality in SU(N) Yang-Mills theory,''
  [arXiv:hep-th/9812090].

\bibitem{Witten}
 E.~Witten,
  ``Some exact multipseudoparticle solutions of classical Yang-Mills  theory,''
  Phys.\ Rev.\ Lett.\  {\bf 38} (1977) 121.
 
\end{thebibliography}
\end{document}